\documentclass[10pt,letterpaper]{article}
\usepackage[top=0.85in,left=2.75in,footskip=0.75in,headheight=43pt]{geometry}

\usepackage{changepage}

\usepackage[utf8]{inputenc}

\usepackage{textcomp,marvosym}

\usepackage{fixltx2e}

\usepackage{amsmath,amssymb}

\usepackage{cite}

\usepackage{nameref,hyperref}

\usepackage{microtype}
\DisableLigatures[f]{encoding = *, family = * }

\usepackage{rotating}

\raggedright
\usepackage[aboveskip=1pt,labelfont=bf,labelsep=period,justification=raggedright,singlelinecheck=off]{caption}

\makeatletter
\renewcommand{\@biblabel}[1]{\quad#1.}
\makeatother

\date{}

\usepackage{lastpage,fancyhdr,graphicx}
\fancyheadoffset[L]{2.25in}
\fancyfootoffset[L]{2.25in}

\begin{document}
\vspace*{0.35in}

\begin{flushleft}
{\Large
\textbf\newline{Markets, herding and response to external information}
}
\newline
\\
Adrián Carro\textsuperscript{*},
Raúl Toral,
Maxi {San Miguel},
\\
Instituto de Física Interdisciplinar y Sistemas Complejos (IFISC), CSIC-UIB, Palma de Mallorca, Spain
\\

%
%





* E-mail: adrian.carro@ifisc.uib-csic.es
\end{flushleft}
\section*{Abstract}
We focus on the influence of external sources of information upon financial markets. In particular, we develop a stochastic agent-based market model characterized by a certain herding behavior as well as allowing traders to be influenced by an external dynamic signal of information. This signal can be interpreted as a time-varying advertising, public perception or rumor, in favor or against one of two possible trading behaviors, thus breaking the symmetry of the system and acting as a continuously varying exogenous shock. As an illustration, we use a well-known German \emph{Indicator of Economic Sentiment} as information input and compare our results with Germany's leading stock market index, the DAX, in order to calibrate some of the model parameters. We study the conditions for the ensemble of agents to more accurately follow the information input signal. The response of the system to the external information is maximal for an intermediate range of values of a market parameter, suggesting the existence of three different market regimes: amplification, precise assimilation and undervaluation of incoming information.



\section*{Introduction}
\label{sec:introduction}


The analysis of financial data has led to the characterization of some non-Gaussian statistical regularities found in financial time series across a wide range of markets, assets and time periods \cite{Mandelbrot1963, Cont2001}. These robust empirical properties are known in the economic literature as \emph{stylized facts}. It has been found, for instance, that the unconditional distribution of returns shows a fat-tailed or leptokurtic character, i.e., it shows a higher concentration of probability in the center and in the tails of the distribution as compared to the Gaussian \cite{Mandelbrot1963}, leading to a higher probability of large returns. A second example is the intermittent behavior of the volatility, measured as absolute or squared returns. This property, known as volatility clustering, implies a tendency for calm and turbulent market periods to cluster together \cite{deVries1994}. This temporal bursting behavior of the volatility leads to the existence of positive autocorrelations for absolute and squared returns which decay only slowly as a function of the time lag \cite{Ding1993}.


Two competing hypotheses have been proposed to explain the origin and ubiquity of stylized facts in financial data. On the one hand, the traditional efficient market hypothesis states that markets are efficient, in the sense that they correctly aggregate all available information, and therefore price changes (returns) fully and instantaneously reflect any new information \cite{Fama1970}. Thus, according to this hypothesis, any particular characteristic of the distribution of returns is a direct consequence of the statistical properties of the news arrival process. On the other hand, recent years have witnessed the development of an alternative approach which might be called the interacting agent hypothesis, as coined in \cite{Alfarano2005}. Indeed, a growing number of contributions based on heterogeneous interacting agents \cite{Kirman1992} have interpreted these stylized facts as the macroscopic outcome of the diversity among the economic actors, and the interplay and connections between them \cite{Kirman1991, Kirman1993, Brock1997, Lux1999, Cont2000}. Heterogeneity refers here to the agents' different level of access to available information or to their capability to choose from a set of various market strategies or trading rules. Regarding their interplay, different interaction mechanisms have been studied, whether direct or indirect, global or local. Thus, according to this hypothesis, any statistical regularity found in financial time series is an emergent property endogenously produced by the internal dynamics of the market.


Along the lines of earlier works \cite{Lux2006}, we can classify the different agent-based financial models into three broad categories. The first one seeks inspiration in well known critical systems from the statistical physics literature, and it is able to reproduce non-Gaussian statistics by carefully adjusting model parameters near criticality \cite{Stauffer1999, Cont2000, Bornholdt2001, Iori2002}. The second group of models, inspired by the seminal work \cite{Brock1997}, assumes that agents interact globally through the price mechanism and public information about the performance of strategies subject to noise \cite{Hommes2006, Chang2007}. These models are able to reproduce some of the aforementioned stylized facts when their signal-to-noise ratio is adjusted around unity. Finally, the third category is composed by a series of stochastic models of information transmission inspired by Kirman \cite{Kirman1991, Kirman1993}, and whose main ingredient is their emphasis on the processes of social interaction among agents, based on herding behavior or a tendency to follow the crowd \cite{Lux1999, Eguiluz2000, Alfarano2005}. These models endogenously give rise to some of the universal statistical properties characterizing financial time series.


If the efficient market approach neglected the importance of the interplay and connections between different economic actors, agent-based finance literature has traditionally overlooked the influence of external sources of information upon markets, treating them as completely closed entities or subject to a constant level of noise. Our research is motivated by the observation that, even if markets are complex systems able to endogenously give rise to non-Gaussian statistical properties, they are by no means closed entities, insensitive to the arrival of external information. Indeed, the recent availability of tools and methods to retrieve and process large corpora of news has allowed empirical financial research to start collecting evidence on how markets react, for instance, to the publication of: news about companies, economic indices, rumors related to the economy, forecasts and recommendations by analysts, news on world events, and financial information by mass media \cite{Hanousek2009, Kiymaz2001, Jegadeesh2006, Arin2008, Alanyali2013}. A more extensive survey of the literature on the influence of external sources of information can be found in \cite{Lillo2015}, while for a more detailed account of the debate about the endogenous and exogenous components of market dynamics we refer to \cite{Johansen2002, Bouchaud2010}. Most of this literature addresses this question from a purely empirical point of view. As a complement to the aforementioned empirical works, the main purpose of this paper is to contribute to the theoretical assessment of the influence of an external source of information upon an agent-based financial market characterized by the existence of a certain herding behavior. Furthermore, we focus on the impact of this exogenous information upon the traders' behavior and how this leads to the existence of different market regimes regarding the input-output response (for other approaches, see \cite{Harras2011, Shapira2014, Golub2015, Rambaldi2015}). In particular, we develop a financial market herding model along the lines of the works by Kirman \cite{Kirman1991, Kirman1993} and Alfarano et al. \cite{Alfarano2008} but open to the arrival of external information affecting the traders' behavior. As an example of information input, we use a well-known \emph{Indicator of Economic Sentiment} published in Germany by the Center for European Economic Research (ZEW). Furthermore, we compare the model results with the main German stock market index, the DAX, in order to find appropriate values for some of the model parameters. The conditions for the ensemble of agents to more accurately follow the information input signal are studied. Interestingly, we find a resonance phenomenon, i.e., a maximum in the response of the system to the external information for an intermediate range of values of a market parameter related to the importance of random behavior relative to herding among traders. This result suggests the existence of different market regimes regarding the assimilation of incoming information.


The outline of the paper is as follows. The first section, describing the methods used, is divided in three subsections. In the first of them we present the original herding model used as a point of departure for our investigation, as well as some analytical results for the sake of comparison in subsequent sections. A brief description of the market framework and the economic variables used is given in the second subsection. The third subdivision contains a detailed discussion about the kind of external information we refer to, a characterization of the particular input signal used as illustration, and the analytical development of a herding model open to external information. The numerical results obtained by simulation of this model are covered in the second section. First, we address the effect on the market of varying the convincing power of the external source of information, comparing with real stock market data to find a suitable value for this convincing power. Then, we quantify the quality of the market model in reflecting the arrival of incoming information and discuss the particular features of the resonance phenomenon found. We draw, in the last section, the main conclusions and add some general remarks.


\section*{Methods}
\label{sec:methods}

\subsection*{The original herding model}
\label{ssec:the_original_herding_model}


Inspired by a series of entomological experiments with ant colonies, Kirman \cite{Kirman1993} proposed a stochastic herding formalism to model decision making among financial agents. In the experiments with ants, entomologists observed the emergence of asymmetric collective behaviors from an apparently symmetric situation: when ants were faced with a choice between two identical food sources, a majority of the population tended to exploit only one of them at a given time, turning its foraging attention to the other source every once in a while. In order to explain this behavior, Kirman developed a stochastic model where the probability for an ant to change its foraging source results from a combination of two mechanism. On the one hand, he postulated the existence of a herding propensity among the ants, i.e., a tendency to follow the crowd, which implies the existence of some kind of interaction among them with information transmission. On the other hand, he also assumed the ants to randomly explore their neighborhood looking for new food sources, so every one of them has an autonomous switching tendency or idiosyncratic behavior, which plays the role of a free will.


This simple herding model was reinterpreted by Kirman in terms of market behavior, by simply replacing an ant's binary choice between food sources by a market agent's choice between two different trading strategies. These different strategies may be related to some particular rules for the formation of the agents' expectations about the future evolution of prices, or result from differences in their interpretation of present and past information. For instance, foreign exchange market traders can adopt different tactics, such as a fundamentalist or a chartist forecast of future exchange rate movements. A further example would be the choice between an optimistic or a pessimistic tendency among the chartist traders. In these examples, the Kirman model would be the decision making mechanism among financial agents, who decide whether to buy or sell in a given situation, thus giving rise to market switches between a dominance of one or the other strategy.


A series of subsequent papers \cite{Lux1999, Alfarano2005, Alfarano2008, Alfarano2009, Alfarano2011} has focused on explaining some of the stylized facts observed in empirical data from financial markets in terms of herding models of the Kirman type. However, there have been two different implementations of the herding term in the literature. In his seminal 1993 paper, Kirman proposed a herding probability that, for each agent, was proportional to the fraction of agents in the opposite state. One of the main drawbacks of this original formalization has been pointed out to be its lack of robustness with respect to an enlargement of the system size $N$, or \mbox{$N$-dependence}, since an increasing number of participants in the market causes the stochasticity to vanish and therefore the stylized facts to fade away. On the contrary, some later authors \cite{Alfarano2005, Alfarano2008, Alfarano2009, Alfarano2011} avoided this problem with an alternative modeling of the interaction mechanism based on a herding probability that, for each agent, is proportional to the absolute number of agents in the opposite state, thus allowing each individual to interact with any other regardless of the system size. We will hereafter adopt this second and more recent formalism. This approach has proven to be successful in reproducing, for instance, the fat tails in the distribution of returns, the volatility clustering, and the positive autocorrelation of absolute and squared returns.


Let us now briefly review the formalization of the Kirman model (in its $N$-independent formulation) and some analytical derivations along the lines of previous works \cite{Alfarano2008, Kononovicius2012} which will be useful for subsequent analyses. Let the market be populated by a fixed number of traders $N$, and let $n$ be the number of those agents choosing one of the two possible strategies, while \mbox{$N - n$} choose the other one. For the sake of clarity, we will hereafter refer to the case of optimistic vs. pessimistic opinions about the future evolution of prices and, in particular, we will call the first group of $n$ agents optimistic and the second group of $N - n$ agents pessimistic. In this manner, \mbox{$n \in {0,1,...,N}$} defines the configuration or state of the system. Its evolution is given by the two aforementioned terms: on the one hand there are pairwise encounters of agents, after which one of them may copy the strategy of the other, and on the other hand there are idiosyncratic random changes of state, playing the role of a free will. An additional assumption of the model is the lack of memory of the agents, so their probability of changing state does not depend on the outcome of previous encounters, neither on previous idiosyncratic switches. Therefore, the stochastic evolution of the system can be formalized as a Markov process depending, at each time step, just on the probability to switch from the present configuration of the system $n$ to some other state $n'$ in a time interval $\Delta t$, denoted by $P(n', t + \Delta t | n, t)$. However, if this time interval $\Delta t$ is taken to be small enough, then the probability to observe multiple jumps is negligible and we can restrict our analysis to $n' = n \pm 1$. Furthermore, the probabilities would then be related to the transition rates per unit time as \mbox{$P(n', t + \Delta t | n, t) = \pi(n \rightarrow n') \Delta t$}. The transition rates for each individual $i$, $\pi^+_i = \pi_i(\textit{pessimistic} \rightarrow \textit{optimistic})$ and $\pi^-_i = \pi_i(\textit{optimistic} \rightarrow \textit{pessimistic})$, can be formally defined as
\begin{displaymath}
	\begin{aligned}
	\pi^+_i  &= a + h \, n,\\[5pt]
	\pi^-_i  &= a + h \, (N - n),
	\end{aligned}
	\end{displaymath}
where the parameters $a$ and $h$ stand for the idiosyncratic switch and the herding intensity coefficients respectively. The rates for the whole system are, therefore,
\begin{equation}
	\begin{aligned}
	\pi^+ (n) &=& \pi(n \rightarrow n + 1) &=& (N - n) \, \left( a + h \, n \right),\\[5pt]
	\pi^- (n) &=& \pi(n \rightarrow n - 1) &=& n \, \left( a + h \, (N - n) \right).
	\end{aligned}
	\label{rates}
\end{equation}
There are two parameters in the model, $a$ and $h$, but one of them can be used as a rescaling of the time variable, so that there is only one relevant parameter, such as $\epsilon = a/h$.


For the sake of subsequent analytical derivations, it is useful to replace the extensive variable $n$, in the range $n \in [0,N]$, by an intensive one which can be treated as continuous for large system sizes $N$ ---note, however, that the limit of an infinite number of agents is never the case in real social and economic systems, where finite-size effects may play a role \cite{Toral2007}---. In particular, we choose $x = 2n/N - 1$ as our intensive variable, giving an opinion index in the range $x \in [-1,+1]$. We choose this range for our intensive variable, as done in \cite{Alfarano2008}, in order to make it comparable to the external information signal which will be used later, also in the range $[-1,+1]$. Note that an opinion index $x=0$ would imply a perfect balance of opinions, while $x=-1$ and $x=+1$ would signal a full agreement on the pessimistic and the optimistic opinions respectively. Note as well that the partial derivation with respect to the new intensive variable becomes, in terms of the previous extensive one, $\partial/\partial x = \left(N/2 \right)\partial/\partial n$ and therefore the relation between their probabilities is $P(x,t) = P(n,t)N/2$.


By means of a systematic and consistent expansion in $N$, it has been shown that the Markovian stochastic process defined above can be approximated by a continuous diffusion process described by the Fokker-Planck equation \cite{Alfarano2008},
\begin{equation}
	\begin{aligned}
	\frac{\partial P(x,t)}{\partial t} &=  \frac{\partial}{\partial x} \Big[ -\mu(x) P(x,t) \Big] + \frac{1}{2}\frac{\partial^2}{\partial x^2} \Big[ D(x) P(x,t) \Big]\\[5pt]
	&= \frac{\partial}{\partial x} \Big[ 2 a x P(x,t) \Big] + \frac{1}{2}\frac{\partial^2}{\partial x^2} \left[ \left( \frac{4 a}{N} + 2 h (1 - x^2) \right) P(x,t) \right],
	\end{aligned}
	\label{FP}
\end{equation}
where $\mu(x)$ plays the role of a drift term and $D(x)$ is the diffusion coefficient. As an alternative way to analyze the dynamics of the model, it is possible to derive a stochastic differential equation for the stochastic process $x(t)$, known as Langevin equation. Using the Fokker-Planck equation~\eqref{FP} and applying the usual transformation rule \cite{vanKampen2007}, within the Itô convention \cite{Ito1951}, we find the Langevin equation describing the process,
\begin{equation}
	\begin{aligned}
	\dot{x} &= \mu(x) + \sqrt{D(x)} \cdot \xi(t)\\[5pt]
	&= - 2 a x + \sqrt{\frac{4 a}{N} + 2 h (1 - x^2)} \cdot \xi(t),
	\end{aligned}
	\label{Langevin}
\end{equation}
where $\xi(t)$ is a Gaussian white noise, i.e., a random variable with a zero mean Gaussian distribution, \mbox{$\langle \xi(t) \rangle = 0$}, and correlations \mbox{$\langle \xi(t) \xi(t') \rangle = \delta(t-t')$}.


Let us first analyze the role played by the noise or diffusion term inside the square root in Eq.~\eqref{Langevin}, \mbox{$D(x) = \frac{4 a}{N} + 2 h (1 - x^2)$}. The first term, dependent on $a$ and inversely proportional to the system size $N$, is related to the ``granularity'' of the system, and so it vanishes in the continuous limit \mbox{$N \to \infty$}. It basically states that for any finite system there are always finite-size stochastic fluctuations related to the fact that the agents have the ability to randomly change their choice. The second term of the diffusion function is a multiplicative noise term, i.e., a noise whose intensity depends on the state variable itself. Furthermore, it is the only term dependent on the herding coefficient, so we will refer to it hereafter as herding term. As this multiplicative noise is maximum for $x=0$ and vanishes for $x= \pm 1$, it is clear from Eq.~\eqref{Langevin} that it tends to move the system away from the center and towards those extremes by making any random partial agreement on one or the other possible opinions grow to a complete consensus.


Regarding the deterministic drift term \mbox{$\mu(x) = -2 a x$}, it is evident form Eq.~\eqref{Langevin} that it has the role of driving the system back to a balanced position at the center of the opinion index, $x=0$. Therefore, we have a competition between two opposed driving forces. One of them is of a stochastic nature, it is dominated by the herding term for large $N$ and it tends to favor the formation of a majority of traders sharing the same opinion about the future evolution of prices. Whereas the other is of a deterministic nature, it is related to the idiosyncratic switches and it tends to break these majorities and drive the system back to a balanced situation, where the traders are equally distributed between both opinions. The magnitude of these two driving forces is related to their respective parameters, $a$ and $h$. Note that in the extreme case of pure herding, $a = 0$, the consensus states $x= \pm 1$ become absorbing: once all the agents agree in the use of a certain strategy, the system is frozen and there is no further evolution [see Eq.~\eqref{Langevin}]. On the opposite, in the extreme case of pure random changes of opinion, $h = 0$, the system would just be characterized by finite-size Gaussian fluctuations around the mean opinion index $\langle x \rangle = 0$.


For non-zero values of both the idiosyncratic and the herding parameters there is a competition between the two aforementioned driving forces and, depending on their relative magnitude, one or the other behavior prevails: either the tendency to form a large majority of agents sharing the same opinion or the tendency to reverse any random majority to a balanced situation. Indeed, the particular functional form of the noise in the Kirman model induces a transition in the dynamics of the system from a monostable to a bistable behavior when increasing the value of $h$ relative to that of $a$. This transition, as well as the particular implications of mono and bistability, can be explained in terms of the probability distribution $P_{\textrm{st}}(x)$, steady state solution of the Fokker-Planck equation~\eqref{FP}, which can be written as
\begin{equation}
	P_{\textrm{st}}(x) = \mathcal{Z}^{-1} \left[ \frac{a}{2Nh} + \frac{(1 - x^2)}{4} \right]^{\frac{a}{h} - 1}.
	\label{solution}
\end{equation}
The normalization factor $\mathcal{Z}^{-1}$ is given by
\begin{equation}
	\mathcal{Z}^{-1} = \frac{\frac{1}{2} \left( \frac{2 \epsilon}{N} + 1 \right)^{-\epsilon} \sqrt{\frac{2 \epsilon}{N} + 1}}{B\left[ \frac{1}{2}\left( 1 + \sqrt{\frac{2 \epsilon}{N} + 1} \right);\, \epsilon,\, \epsilon \right] - B\left[ \frac{1}{2}\left( 1 - \sqrt{\frac{2 \epsilon}{N} + 1} \right);\, \epsilon,\, \epsilon \right]},
\end{equation}
where $B[x;a,b]$ is the incomplete beta function, defined as
\begin{equation}
	B[x;a,b] = \int_0^x u^{a-1}(1-u)^{b-1} du,
\end{equation}
and $\epsilon = a/h$.

Observing the functional form of the steady state solution~\eqref{solution}, one can notice that the sign of the exponent will determine whether the probability distribution is unimodal with a peak centered at $x = 0$ or bimodal with peaks at the extreme values $x = -1$ and $x = +1$. Therefore, when the idiosyncratic switching $a$ is larger than the herding intensity $h$, we find a unimodal distribution, meaning that, at any point in time, the most likely outcome of an observation is to find the community of traders equally split between both options. On the contrary, when the herding $h$ exceeds the idiosyncratic switching intensity $a$, a bimodal distribution is found, meaning that, at any point in time, the most likely outcome of a static observation is to find a large majority of agents choosing the same option. Nevertheless, in different observations, the option chosen by the majority may be different. Note as well that when $a = h$ ($\epsilon = 1$) the probability distribution is uniform, meaning that any share of agents between the two options is equally probable. Because of the ergodicity of the model, these probability distributions can also be understood in terms of the fractional time spent by the system in each state.

\begin{figure}[ht!]

	\centering

	\includegraphics[width=0.7\textwidth, height=!]{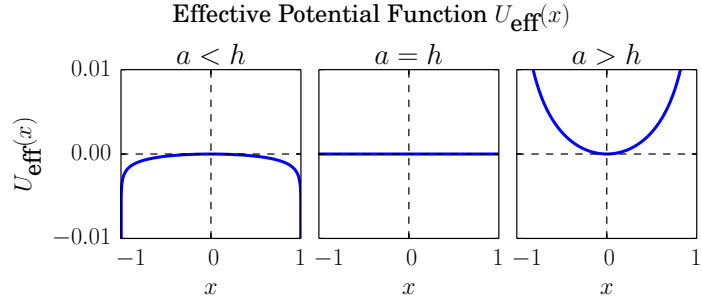}

	\caption{\label{fig1}{\bf Effective potential $U_{\textrm{eff}}(x)$.} For three different values of the idiosyncratic switching tendency, $a = 10^{-4},\,\, 10^{-3},\,\, 10^{-2}$. The rest of the parameter values are $h=10^{-3}$ and $N = 200$. Note that the values of $a$ shown here correspond to the three main cases $a < h$, $a = h$ and $a > h$.}

\end{figure}

An alternative way to observe the transition between a monostable and a bistable behavior and to explain it as a noise induced phenomenon consists in introducing the \emph{effective potential} \cite{SanMiguel2000}: a function combining the effects of both the deterministic driving force and the noise such that its minima are attractive points of the dynamics. We can define the effective potential $U_{\textrm{eff}}(x)$ by assuming an exponential functional form for the stationary probability distribution $P_{\textrm{st}}(x)$,
\begin{equation}
	P_{\textrm{st}}(x) \equiv \mathcal{C}^{-1} \exp \left(-\frac{U_{\textrm{eff}}(x)}{D} \right),
	\label{effPotDef}
\end{equation}
where $D$ is an effective noise intensity that we take as $D=h$, and the constant $\mathcal{C}^{-1}$ plays the role of a normalization factor. Note that, defined as such, the minima of this effective potential function correspond to maxima of the stationary state probability distribution. Let us directly write here the effective potential for the Fokker-Planck equation~\eqref{FP} which takes into account the effect of the multiplicative noise over the deterministic driving force related to the drift term (see \nameref{S1_Appendix} for a derivation),
\begin{equation}
	U_{\textrm{eff}}(x) = (h - a) \, \ln(1-x^2).
	\label{effPot}
\end{equation}
The change of sign occurring in Eq.~\eqref{effPot} for $h = a$ marks the transition from a one well to a double well potential when increasing $h$. The functional form of the effective potential is shown in Fig.~\ref{fig1} for the three possible cases: $a<h$, $a=h$ and $a>h$.

In the $a<h$ case, it is worthy of remark the fact that, although both minima of the effective potential at the extremes of the opinion index space are theoretically infinite wells, in fact, they are not absorbing states and so the system will leave them with probability proportional to $a$. This can be understood by reexamining the Langevin equation~\eqref{Langevin}, where we notice that precisely at the extreme states $x \pm 1$ the only term acting upon the system is the deterministic force driving it towards the center of the opinion index. To sum up, in the $a<h$ case, the role of the herding ($h$) is to induce a bistable effective potential with two wells at the extremes of the opinion index, while that of the idiosyncratic switching ($a$) is to allow for transitions between these two wells.


Two examples of stochastic realizations of this original Kirman model are shown in the two panels in the first row (labeled $F=0$) of Fig.~\ref{fig3} (in the third subdivision of this section). The first example, with an idiosyncratic switching coefficient smaller than the herding intensity ($a = 5 \cdot 10^{-4}$, $h = 10^{-3}$), corresponds to a bimodal probability distribution of states. We can observe, in this panel, the tendency of the system to be temporarily absorbed in the proximity of the consensus states with random switches between them. Some of those switches are not successful and the system returns to the previous consensus before reaching the opposite one. This type of evolution corresponds to a market where traders strongly tend to agree on their opinion about the future evolution of prices but this forecast agreement switches, from time to time, from optimism to pessimism and vice versa. The case of a unimodal probability distribution of states is displayed in the second panel, corresponding to an idiosyncratic switching coefficient larger than the herding intensity ($a = 5 \cdot 10^{-3}$, $h = 10^{-3}$). Note that, as opposed to the previous example, the opinion index spends most of the time in the central part of the opinion space, between $x = -0.5$ and $x = 0.5$. This corresponds to a market where traders are mostly guided by their own idiosyncratic drives, paying little attention to other traders' attitudes, and thus statistically tending to be equally divided among the two possible opinions.


\subsection*{The financial market framework}
\label{ssec:the_financial_market_framework}


In order to model a financial market, we need to embed the stochastic herding formalism described above into an asset pricing framework. Different implementations of the market have been proposed in the literature \cite{Kirman1991, Kirman2002, Lux1999, Alfarano2005, Kononovicius2012}, characterized by different degrees of complexity. We will use here a very simple noise trader framework along the lines of previous works \cite{Alfarano2008}. To this end, we need to define the different types of agents acting in the market and relate them to the herding two-state dynamics. In particular, the market is assumed to be populated by two kinds of agents: fundamentalist and noise traders.


Fundamentalist traders assume the existence of a ``fundamental'' price of the traded asset, towards which the actual market price tends to come back. Therefore, they buy (sell) if the actual market price $p$ is below (above) their perceived fundamental value $p_f$. Assuming that their reaction depends on the log relative difference between the fundamental value and the current market price, instead of absolute difference, the excess demand by the fundamentalist group is given by
\begin{equation}
	ED_f = N_f T_f \ln\left(\frac{p_f}{p}\right),
	\label{EDf}
\end{equation}
where $N_f$ is the number of fundamentalists in the market and $T_f$ is their average traded volume. The reaction to relative rather than absolute under- and overvaluations not only seems more plausible, but also facilitates subsequent derivations. In any case, the small observed daily changes of price ($\sim 1\%$) assure that results would not diverge much if absolute differences were to be used.


The agents of the second group, noise traders, react according to their particular forecast of the future evolution of prices, which can be optimistic or pessimistic. They are therefore divided into two subgroups: optimistic noise traders expect the price of the traded asset to increase in the future and thus decide to buy at the current market price, while pessimistic noise traders expect the price to decrease and thus choose to sell. It is precisely here where the herding model introduced in the previous section enters into the market framework: it is the decision-making mechanism used by the noise traders to choose whether they are optimistic or pessimistic regarding the future price of the traded asset. In this manner, the excess demand by this group of agents becomes a direct consequence of the dominance of optimism or pessimism among them, quantified by the opinion index $x = 2n/N - 1$ introduced above, and can be written as
\begin{equation}
	ED_c = N T_c x,
	\label{EDc}
\end{equation}
where $N$ is the number of noise traders in the market and $T_c$ their average traded volume. Note that only the noise traders $N$ are considered in the herding model described in the previous section.


An equation for the evolution of price can be found by using the Walrasian assumption that relative asset price changes are proportional to the excess demand for the asset \cite{Walras1954}. Commonly referred to as Walrasian \emph{t\^atonnement}, it has become the standard approach in the context of general equilibrium theory \cite{Samuelson1965}. The dynamics of price adjustment can be expressed in continuous-time as
\begin{equation}
	\frac{1}{\beta p} \frac{dp}{dt} = N_f T_f \ln\left(\frac{p_f}{p}\right) + N T_c x,
	\label{pAdjust}
\end{equation}
with $\beta$ representing a price adjustment speed. We further assume, without loss of generality, an instantaneous market clearing ($\beta \to \infty$) and that the total volume traded by both groups of agents is identical ($N_f T_f = NT_c$). We find, in this manner, an equilibrium price driven by both the fundamental value perceived by the fundamentalists and the opinion index among the noise traders,
\begin{equation}
	p(t) = p_f \exp\left( x(t) \right),
	\label{pEq}
\end{equation}
where we have also considered a fundamental value independent of time. This approximation seems plausible in cases where movements of opinion among noise traders occur on a much shorter time scale than changes in the fundamentals of the traded asset, and we are interested in the short time scale behavior. Note that, being the price given by a strictly increasing function of the opinion index among the noise traders, following the majority (herding) is equivalent to following the trends in the price.


For studying the non-stationary properties of returns and volatility, we define the continuously compounded return over an arbitrary time window $\Delta t$ as the logarithmic change of price,
\begin{equation}
	R(t,\Delta t) = \ln p(t+\Delta t) - \ln p(t) = x(t+\Delta t) - x(t),
	\label{return}
\end{equation}
and we use absolute returns as a measure of the volatility, $V(t,\Delta t) = |R(t,\Delta t)|$. For clarity, we will refer hereafter to the daily ($\Delta t = 1~\textrm{day}$) returns and volatility as $R(t)$ and $V(t)$ respectively. Furthermore, for comparison with real data in the Results and Discussion section, we will use the normalized daily returns and volatility, defined as
\begin{equation}
	r(t) = \frac{R(t) - \langle R \rangle}{\sigma(R)}, \qquad \qquad v(t) = |r(t)|,
	\label{normRet}
\end{equation}
where $\langle R \rangle$ and $\sigma(R)$ are, respectively, the mean and the standard deviation of the time series of daily returns. Finally, the autocorrelation of the normalized daily volatility will also be used for comparison with real data:
\begin{equation}
	\textrm{ACF}(v) = \frac{ \Bigl\langle \Bigl(v(t) - \langle v \rangle \Bigr) \Bigl(v(t+\tau) - \langle v \rangle \Bigr) \Bigr\rangle }{ \sigma(v)^2},
	\label{acf}
\end{equation}
where $\langle \cdot \rangle$ denotes an average over time, $\sigma(\cdot)$ stands for the standard deviation, and $\tau$ plays the role of a time lag.

The use of this asset pricing framework with the stochastic herding formalism introduced in the previous subsection gives rise to a market model closed to any external information. The implications of this market model are analyzed in the Results and Discussion section as a particular case of a more general market model open to the arrival of external information (developed in the following subsection): the case of a zero influence external signal.


\subsection*{The model with external information}
\label{ssec:the_model_with_external_information}


The model described so far represents financial markets as completely closed entities, being the price changes of a given asset determined only by the endogenous evolution of the opinion index among noise traders [see Eq.~\eqref{pEq}]. Even if we allow for a time-dependent fundamental value $p_f$, this would only account for the instantaneous arrival of objective information regarding the fundamentals of the asset itself. For instance, the information released in a company's quarterly earnings report directly influences the fundamental value of its stock in the market \cite{Healy2001}. A further example is the instantaneous effect of the devaluation of a given currency on its fundamental value in the exchange market. However, we are not interested here in changes of the perceived fundamental value of an asset resulting from a rational analysis by the fundamentalist traders, and giving rise to direct and linearly proportional movements of the market price [see Eq.~\eqref{pEq}]. On the contrary, we are interested in the arrival of external information not necessarily related to the traded asset and giving rise to trends of optimism or pessimism among the noise traders ---note that, even when dealing with objective information related to the fundamentals of an asset, its disclosure may not only change the value perceived by fundamentalists, but also trigger important speculative movements among noise traders---. Thus, we are concerned with how external news can change the subjective perception or mood of noise traders and generate fads: a prevalence of fear or confidence promoted from outside the market. Note that we focus our attention on a global and passive reception of external information, rather than an individual and active search for it, as studied in \cite{Preis2010, Preis2013, Moat2013, Curme2014}. As examples of this kind of passively received external information we can mention: the publication of news related to companies by specialized financial media \cite{Joulin2008, Alanyali2013, Lillo2015}; the spread of rumors related to the economy \cite{Kiymaz2001}; the updating of various economic indices, such as those tracking the general performance of the economy \cite{Hanousek2009, Rangel2011}; the disclosure of forecasts and recommendations by different analysts \cite{Jegadeesh2006}; the announcement of world events, such as terrorist attacks \cite{Arin2008, Drakos2010}; and, in general, the molding of public opinion by mass media \cite{Davis2006, Tetlock2007}.


For the sake of illustration, we will use hereafter the \emph{Indicator of Economic Sentiment} \cite{zew}, developed by the Center for European Economic Research (ZEW), as an example of external information input to the market. This indicator measures the level of confidence that a group of up to 350 financial and economic analysts (experts from the finance, research and economic departments as well as traders, fund managers and investment consultants) has about the current economic situation in Germany and its expected development for the next 6 months. The survey is conducted every month and the corresponding index is constructed as the difference between the percentage share of analysts who are optimistic and the percentage share of analysts who are pessimistic about the development of the economy. We nevertheless underline that the formalism that follows is general and independent of the particular shape of the external information signal used. The only relevant features of this information input having a significant effect on the results are its strength and its frequency or rate of change.


In order to design a financial market model open to the arrival of external information of the aforementioned type, the immediate question becomes how to modify the transition rates~\eqref{rates} to take this external input into account. We are here interested in the modification of the social processes of opinion formation and propagation of information among the economic agents. So we are naturally led to introduce the information input signal in the social term of the transition rates, that is, in the herding coefficient $h$ (for a different approach, see \cite{Harras2012}). Note that this choice leads to a direct linear dependence of the effect of the external information upon a given agent on the number of agents with the contrary opinion [see Eqs.~\eqref{timeRates} and~\eqref{herding}]. In particular, this effect completely vanishes when there is no agent in the opposite state. In this manner, optimist (pessimist) traders in a market with a clear consensus for optimism (pessimism) will be less affected by external information in the opposite sense. Thus, we modify the transition rates~\eqref{rates} as\begin{equation}
	\begin{aligned}
		\pi^+ (n,t) &=& \pi(n \rightarrow n + 1,t) &=& (N - n) \left( a + h_+(t) n \right),\\[5pt]
		\pi^- (n,t) &=& \pi(n \rightarrow n - 1,t) &=& n \left( a + h_-(t) (N - n) \right),
	\end{aligned}
	\label{timeRates}
\end{equation}
where the herding coefficients are now different in the two possible directions and are both time-dependent functions given by
\begin{equation}
	\begin{aligned}
		h_+(t) &=& h_0 + \frac{F}{N} i(t),\\[5pt]
		h_-(t) &=& h_0 - \frac{F}{N} i(t),
	\end{aligned}
	\label{herding}
\end{equation}
with $h_0$ playing the role of a constant or background herding coefficient, $F$ acting as the strength or intensity of the external information applied to the whole system, and $i(t)$ being the dynamic information itself. Note that we will refer to $a$ and $h_0$ as parameters of the market and to $F$ as a parameter of the input signal. As the opinion index $x$, the information function $i(t)$ takes values in the range $[-1,1]$, being the negative and positive ones respectively associated with pessimistic and optimistic news. The intensity can be understood as a measure of the resources used by the external source in order to transmit the information and persuade the agents. Note that this information input term is not proportional to the total strength exerted on the system, but to the total strength per agent, $F/N$. The rationale behind this particular functional form for the external input term is basically a limited resources assumption: the resources spent in transmitting the information and convincing the whole system are divided among its constituents, so that if the system size $N$ increases, the resources available for convincing each of the agents decrease. The reason for adding and subtracting the external input term, respectively in $h_+(t)$ and $h_-(t)$, is just so that positive (negative) values of the information help transitions towards optimism (pessimism) while they hinder transitions towards pessimism (optimism). In order to keep the transition rates always positive, the intensity per agent must satisfy $F/N \leq h_0$.


Proceeding in a similar manner as for the original herding model in the first subdivision of this section, and applying the same approximations, we find the new Fokker-Planck equation:
\begin{equation}
	\begin{aligned}
		\frac{\partial P(x,t)}{\partial t}  &=  \frac{\partial}{\partial x} \Big[ -\mu(x,t) P(x,t) \Big] + \frac{1}{2}\frac{\partial^2}{\partial x^2} \Big[ D(x) P(x,t) \Big]\\[5pt]
		&= \frac{\partial}{\partial x} \Big[ \Big( 2 a x - F (1 - x^2) i(t) \Big) P(x,t) \Big] + \frac{1}{2}\frac{\partial^2}{\partial x^2} \left[ \left( \frac{4 a}{N} + 2 h_0 (1 - x^2) \right) P(x,t) \right].
	\end{aligned}
	\label{FP2}
\end{equation}
Note that, compared to the previous Fokker-Planck equation~\eqref{FP}, the herding coefficient $h$ has been replaced by its constant part $h_0$ inside the diffusion coefficient $D(x)$. There is also a new time-dependent term inside the drift function $\mu(x,t)$, which becomes itself dependent on time. Again, the conventional transformation rule leads us, within the Itô form, to the Langevin equation describing the process,
\begin{equation}
	\begin{aligned}
		\dot{x} &= \mu(x,t) + \sqrt{D(x)} \cdot \xi(t)\\[5pt]
		&= - 2 a x + F (1 - x^2) i(t) + \sqrt{\frac{4 a}{N} + 2 h_0 (1 - x^2)} \cdot \xi(t),
	\end{aligned}
	\label{Langevin2}
\end{equation}
where $\xi(t)$ is, as before, a Gaussian white noise.


The constant part of the herding coefficient $h_0$ plays exactly the same role as the herding coefficient itself in the original Kirman model. The new term of the drift function, on the contrary, changes fundamentally the general behavior of the system. In particular, it will force the market to follow the external information signal by favoring or opposing, depending on the sign of this signal, the tendency towards $x=0$ caused by the first drift term. In other words, the equilibrium point of the drift function is no longer a constant at $x=0$, but a function dependent on time through $i(t)$ and taking values around this central point. The new parameter $F$, the strength of the information input, modulates the intensity of this effect. Note also that the factor $(1 - x^2)$ inside the new drift term causes its effects to vanish at the extremes of the opinion index space and its absolute value to be maximal at its center for $i(t) = \pm 1$. Thus, the effect of the external information upon the market vanishes for increasing consensus among the noise traders and becomes strongest when the group is equally divided between the two possible opinions. This behavior seems plausible from the perspective that groups with consensus tend to be confident about their common decision and less prone to pay attention to external sources of information than groups with a division of opinions \cite{Granovetter1978}. Note as well that the new drift term vanishing at the extremes of the opinion index space, it does not help the system to exit the consensus states, and therefore some idiosyncratic behavior ($a>0$) is still needed in order to observe transitions between both full agreement states.


Concerning the competition between the deterministic and the stochastic terms of Eq.~\eqref{Langevin}, the inclusion of an external information input in Eq.~\eqref{Langevin2} has the general effect of counteracting or enhancing the deterministic driving force depending on the sign of this information signal. For a deeper understanding of the transition induced by the multiplicative noise upon the deterministic driving force and the symmetry breaking role of the external information, let us write the effective potential [see Eq.~\eqref{effPotDef}] for the Fokker-Planck equation~\eqref{FP2},
\begin{equation}
	U_{\textrm{eff}}(x,t) = (h_0 - a) \, \ln(1-x^2) - x \, F \, i(t),
	\label{forcedEffPot}
\end{equation}
Again, we leave its derivation for \nameref{S1_Appendix}. Note that the new term, related with the information input signal, is linear in the opinion index variable $x$, and thereby it breaks the symmetry ($x \leftrightarrow -x$) of the potential for $i(t) \neq 0$.


\begin{figure}[ht!]

	\centering

	\includegraphics[width=\textwidth, height=!]{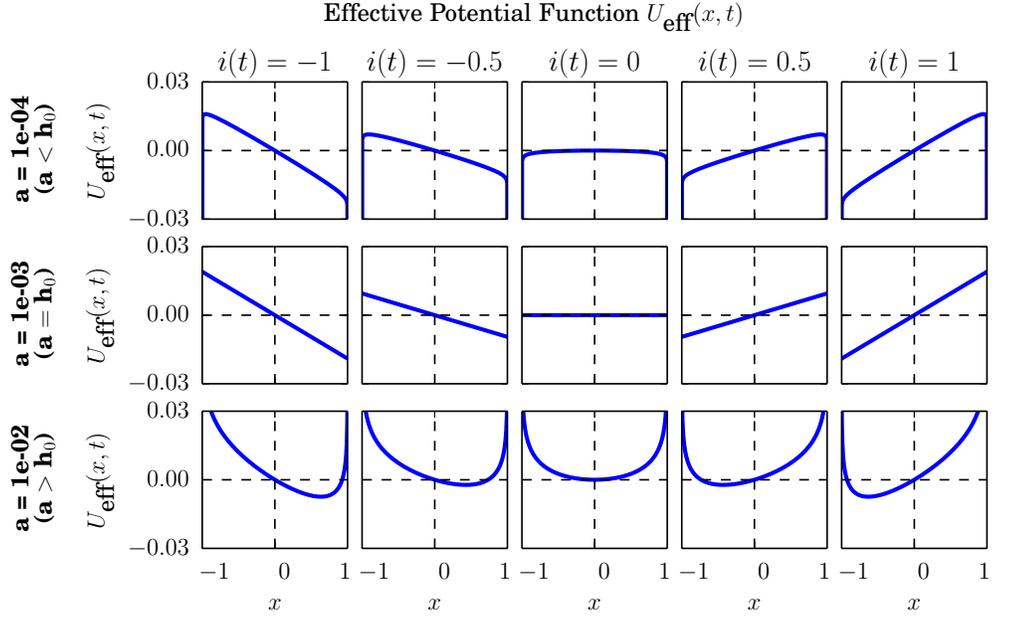}

	\caption{\label{fig2}{\bf Effective potential $U_{\textrm{eff}}(x, t)$.} Snapshots for various values of the information input signal $i(t)$ and for three different values of the idiosyncratic switching tendency, $a = 10^{-4},\,\, 10^{-3},\,\, 10^{-2}$. The rest of the parameter values are $h_0=10^{-3}$, $F = 0.02$, and $N = 200$. Note that the values of $a$ shown here correspond to the three main cases $a < h_0$, $a = h_0$ and $a > h_0$.}

\end{figure}

The dependence of the effective potential on the information input signal is illustrated in Fig.~\ref{fig2}, where snapshots are presented for five different values of the signal and for three values of the idiosyncratic parameter $a$. For values of $a$ below $h_0$, allowing for the creation of a double well effective potential, the role of the external information is to modify the depth of these wells, making one of them relatively more attractive than the other. In this case, large majorities of traders sharing the same opinion tend to emerge in the market, generally including the whole of it, and the external information simply facilitates an optimistic or pessimistic consensus depending on its sign. When $a$ equals $h_0$ the effective potential becomes a linear function, and the role of the information input is to vary its slope, thus creating a unique minimum at $x=-1$ or $x=1$. Therefore, in a market where every share of opinions is equally probable, the external information facilitates again the creation of strong majorities, tending to include the whole of the market. Values of $a$ larger than $h_0$ give rise to a monostable effective potential, where the minimum is moved around the center of the opinion index space by the influence of the information input. Thus, when traders tend to be equally divided between the two possible opinions, the role of the external information is to slightly break this symmetry, giving rise to weak majorities tending to not include the whole of the market.


\section*{Results and Discussion}
\label{sec:results_and_discussion}


We present in this section the main results obtained by numerical analysis of the market model with arrival of external information described in the last part of the previous section. We first analyze the effect of different intensities of the external information upon the market, as well as compare the corresponding results with real financial data in order to find an appropriate value for this new parameter. Once the intensity has been fixed, we study the market conditions for the ensemble of traders to more accurately follow the information input signal, finding an interesting resonance phenomenon: a maximum of the accuracy of the market in reflecting the arrival of external information. This suggests the existence of different market regimes regarding the assimilation of incoming information.


In contrast with the methods presented in the previous literature \cite{Alfarano2005, Alfarano2008, Alfarano2009, Kononovicius2012}, we have used a Gillespie algorithm for the simulation of the model \cite{Gillespie1977, Gillespie1992}. A single realization of this algorithm represents a random walk with the exact probability distribution of the master equation, therefore generating statistically unbiased trajectories of the stochastic equation. Thereby, we generate an unbiased sequence of times when the transitions of agents between the optimistic and pessimistic states take place. For a stochastic system with time-dependent transition rates \mbox{$\pi^{\pm}(t) = \pi(n \rightarrow n \pm 1, t)$}, and assuming that the last transition took place at time $t_1$, the cumulative probability of observing an event $n \rightarrow n \pm 1$ at time $t_2^{\pm}$ can be written as
\begin{equation}
	F(t_2^{\pm}|\,t_1) = 1 - e^{-\int\limits_{t_1}^{t_2^{\pm}} \pi^{\pm} (t) dt}.
\end{equation}
We can equate this expression to a uniformly distributed random number $u^{\pm}$ between $0$ and $1$ in order to find an equation for the time of the next event, $t_2^{\pm}$,
\begin{equation}
	\int\limits_{t_1}^{t_2^{\pm}} \pi^{\pm} (t) dt = -\ln(1-u^{\pm}) \equiv -\ln(u^{\pm}),
	\label{integrationTimes}
\end{equation}
where we have used that $u^{\pm}$ and $1-u^{\pm}$ are statistically equivalent. By that means, two different times are found: $t_2^+$, corresponding to the transition rate $\pi^+ (t)$, and $t_2^-$, related to the transition rate $\pi^- (t)$. The event actually taking place is the one related to the shortest time.


For the sake of solving Eq.~\eqref{integrationTimes}, we assume that the information signal $i(t)$ [and thus the transition rates, see Eqs.~\eqref{timeRates} and~\eqref{herding}], stays constant between every two releases of information. Moreover, we assume without loss of generality that these announcements or news arrival instants are periodical in time. This is particularly the case when dealing with the publication of some economic surveys such as the ZEW \emph{Indicator of Economic Sentiment} ---the input signal we chose for illustration---, which is released monthly. In order to both correctly introduce the external information into the model and compare its results to real data from stock markets, we need to set the relation between the time unit of the model and the real time. For simplicity, we choose the time unit of the model to correspond to a real day of trade. Again, this choice implies no loss of generality, since the time scale of the model ---that is, the velocity at which noise traders change their position in the market--- can also be varied by modifying the values of the parameters $a$ and $h$ while keeping their relation constant. As a consequence, we update the \emph{Indicator of Economic Sentiment} every $20$ time units of the model, corresponding to the $20$ trading days of each month (considering, for simplicity, months of four weeks and weeks of five trading days). All simulations start from a random distribution of optimistic/pessimistic opinions among noise traders and evolve for $5280$ time units, roughly corresponding to the trading days between December of 1991 and November of 2013: the data period of the ZEW \emph{Indicator of Economic Sentiment} that we use. For comparison with real data we use the daily variations of the German stock exchange index DAX during the same period of time. Note that the monthly variations of these two datasets have a small but positive cross-correlation, showing that there is no direct cause-effect relationship between them, but rather that the ZEW \emph{Indicator of Economic Sentiment} constitutes a relevant input to be fed into the model presented above, whose agents will then filter it in a non-trivial and non-linear way through their idiosyncratic changes and their herding interactions.


\subsection*{Effect of the external information on the market}
\label{ssec:effect_of_the_external_information_on_the_market}


The particular modifications of the collective behavior of the market due to the introduction of an external information signal depend on the specific values of the model parameters. We devote this subsection to the analysis of the effect produced by different input signal intensities on the typical simulated patterns of three market variables: the opinion index among noise traders, the normalized daily returns and their bursting behavior, and the autocorrelation function of the daily volatility. For the two latter cases we offer as well a comparison with real financial data (DAX), allowing us to choose an appropriate value for the intensity of the information signal.

\begin{figure}[ht!]

	\centering

	\includegraphics[width=\textwidth, height=!]{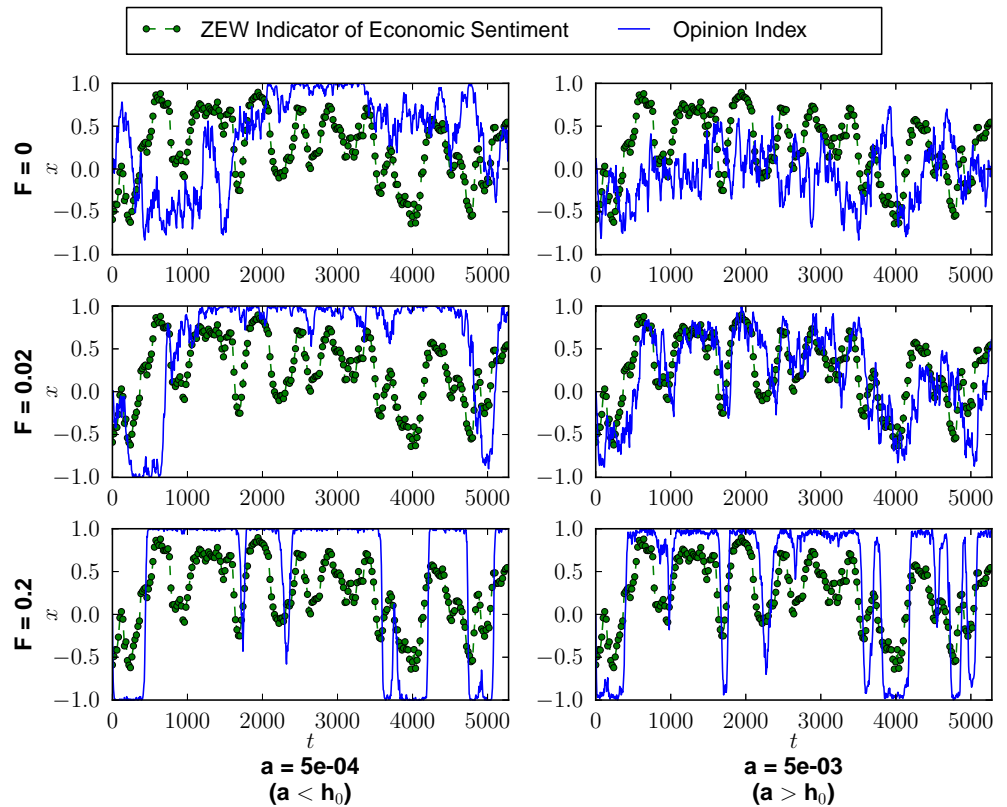}

	\caption{\label{fig3}{\bf Effect of the information intensity on the opinion index.} Green points and dashed green line: External information signal, ZEW \emph{Indicator of Economic Sentiment}, data from December 1991 to November 2013. Solid blue line: Time evolution of the opinion index simulated for different values of the external information intensity $F$ and the idiosyncratic switching tendency $a$. The herding parameter and the system size are fixed as $h_0 = 10^{-3}$ and $N = 200$.}

\end{figure}

Figure~\ref{fig3} contains two sets of three panels illustrating the effect on the noise traders opinion index of increasing the intensity of the incoming information from the closed market case, $F=0$, to its maximum allowed value, \mbox{$F = N h_0$}. The three panels in the first column address the case of market parameters in the bistable regime, $a<h_0$, while the three panels in the second column deal with market parameters in the monostable regime, $a>h_0$. In the first case, $a=5 \cdot 10^{-4}$ and $h_0 = 10^{-3}$, we observe that the application of an information input reinforces the bistability of the distribution of states. By comparing the closed market case ($F=0$) with the market subject to a small information strength ($F=0.02$), we notice that the introduction of a small input intensity is able to modify the behavior of the system by pushing the opinion index towards a fully optimistic or pessimistic extreme. However, the market is not able to follow the mood changes of the external signal: the opinion index may stay around an optimistic extreme while the external information is rather pessimistic (see, for example, the negative peak around $t=4000$). In the maximum information strength case ($F=0.2$), we observe an even faster collapse of the opinion index around its extreme values, but we notice now that the changes of mood of the external signal are generally matched by large opinion movements of the market in the same direction.

The effect of increasing the intensity of the incoming information in the case of market parameters ($a$, $h_0$) in the monostable regime is shown in the second column of Fig.~\ref{fig3}, where $a=5 \cdot 10^{-3}$ and $h_0 = 10^{-3}$. First, we notice that the application of an information input can result in a bistable-like behavior, as the one expected for $a < h_0$ and $F = 0$, especially in the case of a large input intensity (case $F=0.2$). This is due to the introduction of the external information term as part of the herding coefficient [see Eq.~\ref{herding}]. Interestingly, we also notice that for a market with this level of idiosyncrasy even a small input intensity ($F=0.02$) is able to force the opinion index to follow the shape of the external signal. Similarly to the bistable case ($a<h_0$ column), a large input strength ($F=0.2$) compels the market to an amplification of the information signal. Nonetheless, the higher level of idiosyncrasy allows now for the collapse of the opinion index around the fully optimistic or pessimistic extremes to take place as soon as the signal becomes, respectively, optimistic [$i(t)>0$] or pessimistic [$i(t)<0$]. The convincing power of the external information source being so strong, most of the noise traders are quickly persuaded to align their opinions with the optimism or pessimism of the input signal.

\begin{figure}[ht!]

	\centering

	\includegraphics[width=\textwidth, height=!]{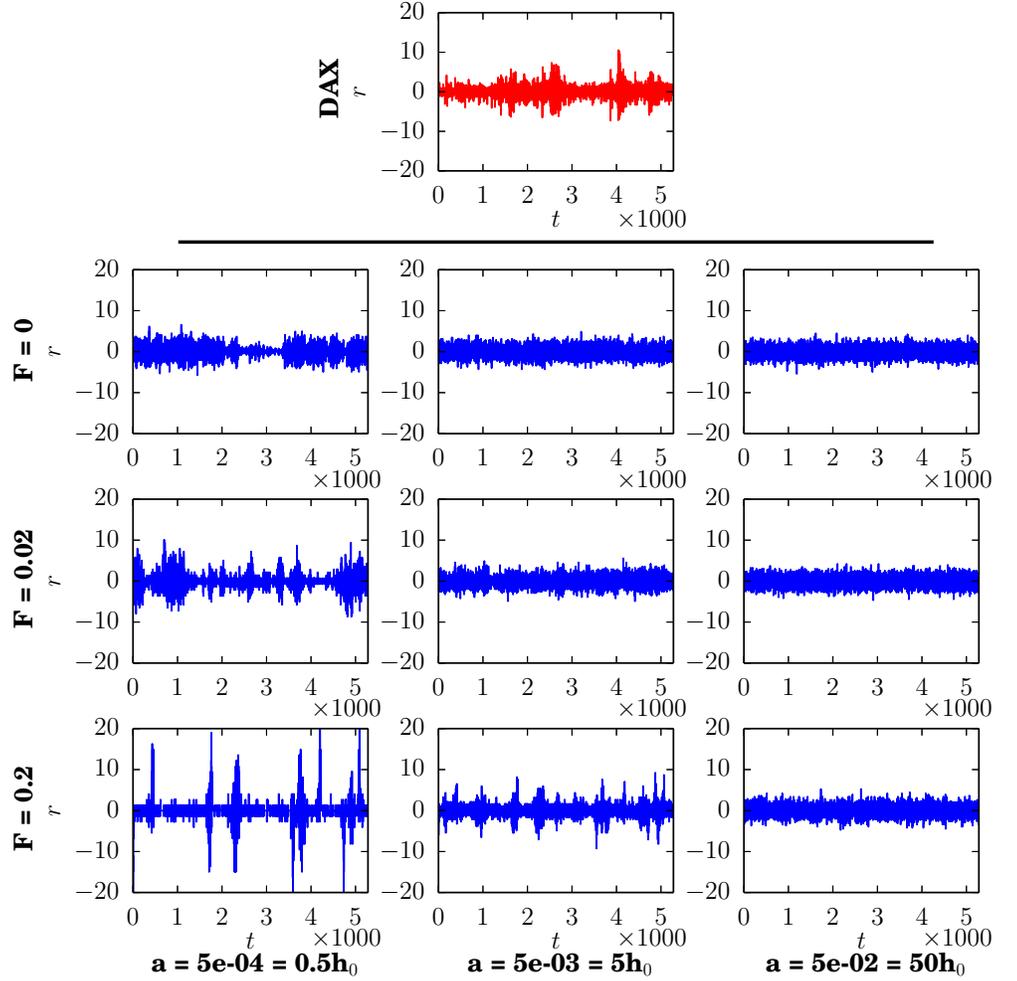}

	\caption{\label{fig4}{\bf Normalized daily returns $r(t)$.} First panel, red line: German DAX index from December 1991 to November 2013, shown for comparison. Table of nine panels, blue lines: Model results for three different input intensities of the external information $F$ and three values of the idiosyncratic switching tendency $a$. The rest of the parameters are fixed as $h_0 = 10^{-3}$ and $N = 200$.}

\end{figure}

The behavior of the normalized daily returns [see Eq.~\eqref{normRet}] under different input strengths and for various market parameters is depicted in Fig.~\ref{fig4}, where the returns of the German DAX index are also displayed in a first panel for visual comparison. Observing the model results for market parameters in the bistable regime (first column, $a=5 \cdot 10^{-4}$, $h_0 = 10^{-3}$), we notice that an evident effect of an increasing convincing power of the external information source is the strengthening of the volatility clustering phenomenon, i.e., the clustering of periods of large returns and periods of small returns. Even if some volatility clustering is already present ---endogenously produced--- in the closed market case ($F=0$), this feature seems to be underrepresented for these market parameters when compared to the DAX data. A clear enhancement of the clustering effect is observed when a low intensity information signal is introduced ($F=0.02$), bringing the model results closer to the DAX data. A large input strength ($F=0.2$), however, results in an unrealistic amplification of the clustering: almost all the returns in the 22 years simulated are realized in less than ten short bursts, when the information input changes sign and the whole market switches from optimism to pessimism or vice versa. A similar but weaker influence of the external information is found for $a=5 \cdot 10^{-3}$ (second column), a value corresponding to the monostable regime but in the same order of magnitude as $h_0$. For $a=5 \cdot 10^{-2}$ (third column), however, the effect of the input signal appears to be negligible regardless of its strength. This can be understood bearing in mind that, being the idiosyncratic switching tendency much larger than the herding propensity, the dynamics is dominated by random changes of opinion of the traders, and therefore the external information is irrelevant in the scale of days, the one used for computing the returns.

\begin{figure}[ht!]

	\centering

	\includegraphics[width=\textwidth, height=!]{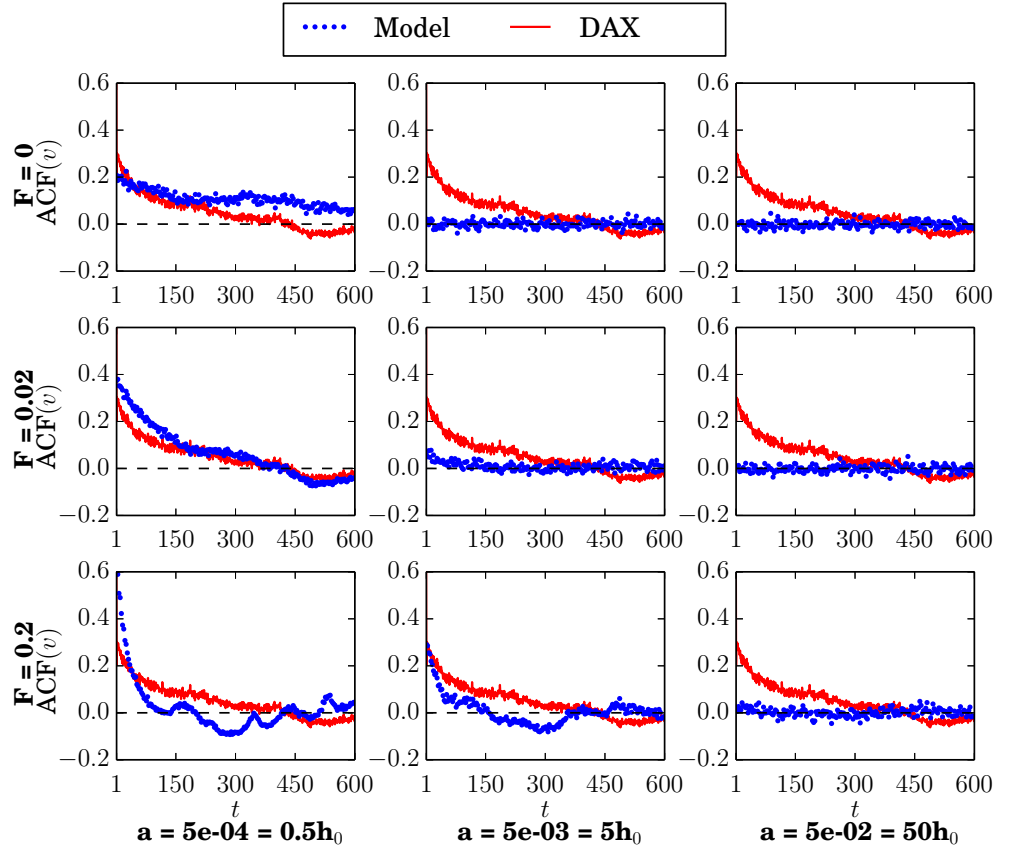}

	\caption{\label{fig5}{\bf Autocorrelation function (ACF) of the normalized daily volatility $v(t)$.} Black dots: Model results for three different input intensities of the external information $F$ and three values of the idiosyncratic switching tendency $a$. The rest of the parameters are fixed as $h_0 = 10^{-3}$ and $N = 200$. Solid blue line: The corresponding ACF of the German DAX index volatility is shown in each case for comparison. The DAX data corresponds again to the period from December 1991 to November 2013.}

\end{figure}

The observation of the time series of the normalized daily returns (Fig.~\ref{fig4}) might lead to think that the parameters used are degenerate, i.e., that different combinations of their values can lead to similar results, as it is the case for the couples $F=0.02$, $a=5 \cdot 10^{-4}$ and $F=0.2$, $a=5 \cdot 10^{-3}$. However, this degeneracy is only apparent, as it can be shown by simply analyzing other magnitudes, for example: the autocorrelation function of the normalized daily volatility [see Eq.~\eqref{acf}], measured as the absolute value of the normalized daily returns, illustrated in Fig.~\ref{fig5}. The same input intensities and market parameters as in the previous figure are studied, and the corresponding autocorrelation function for the DAX normalized daily volatility is shown for comparison in every panel. If we first focus on the examples for market parameters in the bistable regime (first column, $a=5 \cdot 10^{-4}$, $h_0 = 10^{-3}$), we find a highly significant autocorrelation of absolute returns which only falls off slowly, in accordance with previous empirical literature \cite{Ding1993, Mandelbrot1997, Cont2001, Cont2005}. However, in the closed market case ($F=0$) this decrease is extremely slow for large time lags, where we still find a significant autocorrelation, as opposed to the DAX data. The introduction of a small input strength signal ($F=0.02$) is able to modify this behavior, the autocorrelation for large time lags becoming negligible or even slightly negative, as it is found for the DAX index. On the opposite, when the strength of the input is increased up to its maximum value ($F=0.2$), the outcome of the model becomes strongly driven by the shape of the information signal and, as a result, the autocorrelation function becomes as well a direct consequence of this signal shape and very different from the DAX data. Thus, the expected behavior of the autocorrelation of absolute returns for long time lags is found for markets subject to a low intensity information entrance, while it is not present for closed or completely driven markets. For a market with an idiosyncratic tendency larger but in the same order of magnitude as the herding propensity (second column, $a=5 \cdot 10^{-3}$), we observe, on the one hand, a complete lack of autocorrelation for both the closed market example ($F=0$) and the case of a small information influence ($F=0.02$). On the other hand, the entrance of an information signal with its maximum convincing power ($F=0.2$) leads again to an unrealistic behavior of the autocorrelation function, which becomes driven by the shape of the incoming information signal. In the case of larger values of the idiosyncratic switching tendency, and being the market dominated by random opinion changes, we find no significant autocorrelation of absolute returns regardless of the information strength applied.

Note that, with the small value of the information strength, $F=0.02$, and for some values of the market parameters, the model described here is able to reproduce the main statistical features of the DAX index. On the one hand, the model emulates the behavior of the DAX normalized daily returns (see Fig.~\ref{fig4}), giving rise to a comparable volatility clustering effect. On the other hand, the model leads to a similar autocorrelation function of the normalized daily volatility (see Fig.~\ref{fig5}), reproducing both the slow decay of the DAX autocorrelation and its zero and slightly negative values for very long time lags. Note that this last feature, the long time lag behavior of the autocorrelation function, is not captured by the market model closed to external information. For completeness, we also include the probability distributions of absolute normalized daily returns in \nameref{S1_Figure}, for the same parameter values used in Fig.~\ref{fig4} and~\ref{fig5}: while their lack of temporal structure hides any volatility clustering effect, these distributions seem to support our choice of parameter value. In view of these results, we select the small value of the external information intensity, $F=0.02$, to be used in the rest of the research presented hereafter.


\subsection*{Resonance phenomenon}
\label{ssec:resonance_phenomenon}


We have analyzed above the effects caused by different information input intensities on the market, finding that a small strength input ($F = 0.02$) produces results consistent with real financial data. Let us now focus on this case, keeping the external information intensity fixed as $F = 0.02$, and search for the values of the model parameters for which the ensemble of agents follows more accurately the shape of this signal. Thus, we are concerned here with the study of the conditions under which the market best reflects the arrival of external information. As mentioned above, although the discussed market model has in principle three parameters ($a$, $h_0$, and $F$), one of them can be used as a rescaling of the time variable, so that there are only two relevant parameters. Given that the effective influence of the external information upon the system is determined by the relative importance of the input strength $F$ and the background herding coefficient $h_0$ [see Eqs.~\eqref{timeRates} and~\eqref{herding}], we choose to keep the latter fixed and therefore use the idiosyncratic switching tendency $a$ as our control parameter. In particular, we choose the values $h_0 = 10^{-3}$, $F = 0.02$ and $N=200$. The input intensity per agent, $F/N$, is thus ten times smaller than the herding coefficient, its maximum allowed value. We have also performed simulations with different system sizes ($N = 50$ and $N = 800$), finding a generally equivalent behavior which will be discussed below.

\begin{figure}[ht!]

	\centering

	\includegraphics[width=0.85\textwidth, height=!]{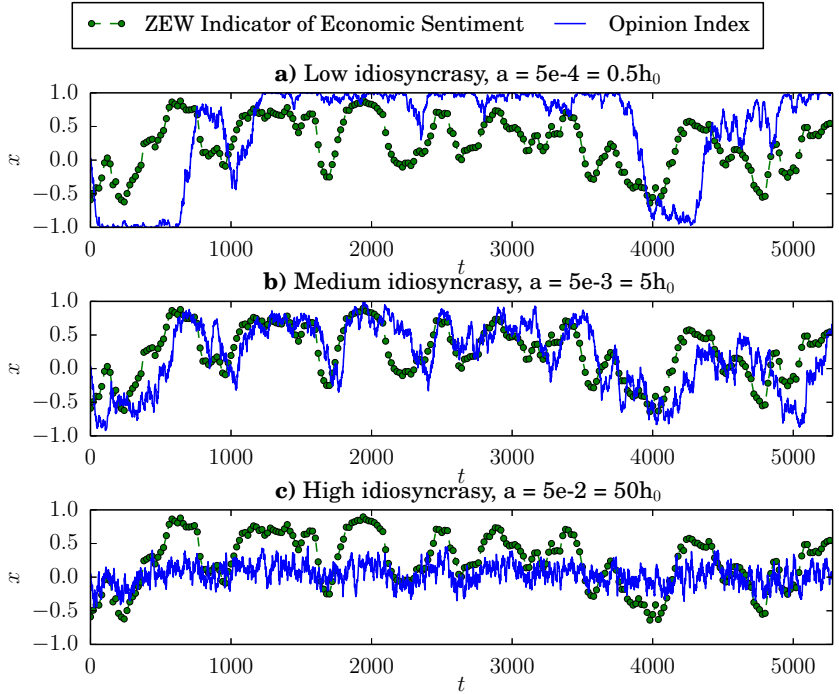}

	\caption{\label{fig6}{\bf Effect of the idiosyncratic switching tendency on the opinion index.} Green points and dashed green line: External information signal, ZEW \emph{Indicator of Economic Sentiment}, data from December 1991 to November 2013. Solid blue line: Time evolution of the opinion index simulated for different values of the idiosyncratic switching tendency $a$ and fixed values of the parameters $h_0 = 10^{-3}$, $F = 0.02$, and $N = 200$.}

\end{figure}

Let us start by considering the influence that varying the idiosyncratic switching tendency $a$ has over the time series of the noise traders opinion index $x$, illustrated in Fig.~\ref{fig6}. Note that there is no maximum allowed value for this parameter, as it was the case with the signal intensity $F$: its only constraint is that it must be $a > 0$ so that the extremes of the opinion index space are not absorbing states. Therefore, we simply choose a reasonable range which includes the different behaviors described in the previous sections and observed in Fig.~\ref{fig2}: from a fully bimodal case ($a \ll h_0$), with almost two deltas at the extremes of the probability distribution of states; up to a fully unimodal case ($a \gg h_0$), with an almost perfect Gaussian distribution of states.

In the first of these cases ($a = 0.5h_0$, panel~\textbf{a}) the system is clearly unable to follow the shape of the input signal and, in fact, the opinion index stays close to a full agreement state for most of the time. This is due to the extremely low level of idiosyncratic behavior relative to the herding tendency: noise traders tend to form a large consensus group which cannot easily be convinced by the external source. Even if a double well effective potential has been induced, the probability of observing a transition between both wells, proportional to $a$, is too small to allow for the group to leave the consensus states at a rate large enough for the market to adapt to the updates of the external information. In the intermediate case ($a = 5h_0$, panel~\textbf{b}), when the idiosyncratic coefficient is slightly larger than the herding tendency, we observe that the system easily follows the shape of the input signal. This can be understood taking into account that intermediate values of $a$ larger than $h_0$ give rise to a rather wide monostable effective potential where the system can still be largely driven by the external signal. Thus, the market seems to reflect rather precisely the arrival of external information. On the opposite, for very large values of the idiosyncratic coefficient ($a = 50h_0$, panel~\textbf{c}) the system is again unable to fit the shape of the input signal and the trajectories of the opinion index seem rather noisy. As shown in the last part of the Methods section, very large values of $a$ lead to narrow unimodal effective potentials with minima moving closely around the center of the opinion index space. Thus, the external information input has an almost negligible influence and the market seems to be unaware of it.


In this way, we note that different values of the idiosyncratic coefficient lead to different levels of coincidence between the opinion index resulting from the simulation of the model and the information signal used as an input. In order to quantify this phenomenon ---that is, in order to measure the quality of the market response in following the external information driving--- we use the input-output correlation ($\textrm{IOC}$), defined as the maximum of the cross-correlation function between the input signal $i(t)$ and the system output $x(t)$,
\begin{equation}
	\textrm{IOC} = \frac{ \max_{\tau} \Bigl\{ \Bigl\langle \Bigl(i(t) - \langle i \rangle \Bigr) \Bigl(x(t+\tau) - \langle x \rangle \Bigr) \Bigr\rangle \Bigr\} }{\sigma(i) \sigma(x)},
	\label{ioc}
\end{equation}
where $\langle \cdot \rangle$ denotes an average over time, $\sigma(\cdot)$ stands for the standard deviation, $\tau$ plays the role of a time lag, and $\max_{\tau} \{ \cdot \}$ finds the maximum value of a $\tau$-dependent function \cite{Collins1996}. Note that $\textrm{IOC}$ is a scalar measure quantifying the maximum of the input-output cross-correlation function, which depends on the time lag $\tau$. A larger $\textrm{IOC}$ is related with a better entrainment of the market by the external information signal, corresponding its maximum value, $\textrm{IOC} = 1$, to a perfect fit between the time series $i(t)$ and $x(t)$. Note that an amplification of the input signal is understood here as a worse fit when compared with a perfect input-output correspondence.

\begin{figure}[ht!]

	\centering

	\includegraphics[width=0.85\textwidth, height=!]{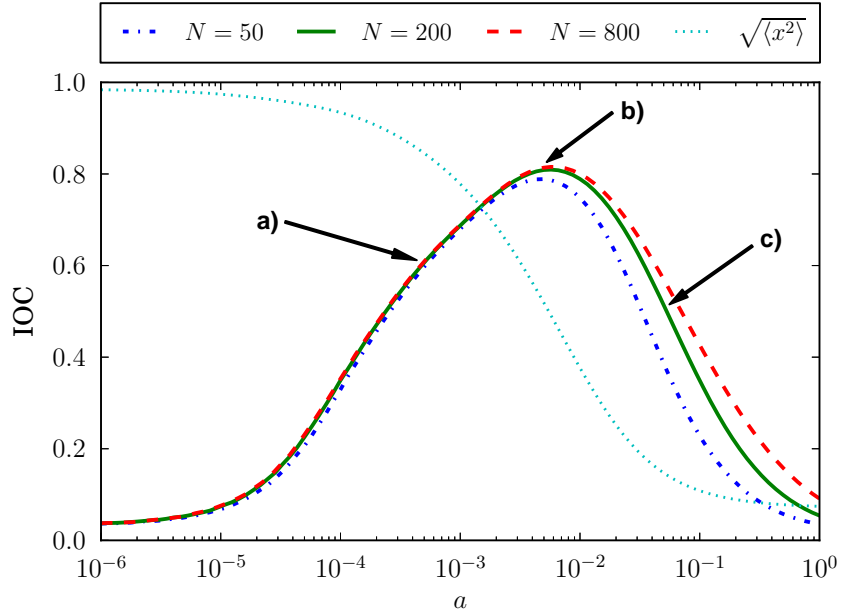}

	\caption{\label{fig7}{\bf Input-output correlation $\textrm{IOC}$ as a function of the idiosyncratic switching tendency $a$.} The correlation is computed according to Eq.~\eqref{ioc}. Parameter values are $h_0 = 10^{-3}$, $F = 0.02$, and three different system sizes, $N = 50,\,\, 200,\,\, 800$. A measure of the fraction of time spent by the system near the extremes of the opinion index space, $\sqrt{\langle x^2 \rangle}$, is also shown for the $N = 200$ case. Arrows point at the idiosyncrasy levels whose opinion index time series are shown in Fig.~\ref{fig6}, marked also by their corresponding letter: a)~$a = 5 \cdot 10^{-4}$, b)~$a = 5 \cdot 10^{-3}$, c)~$a = 5 \cdot 10^{-2}$}

\end{figure}

The results obtained for the input-output correlation are displayed in Fig.~\ref{fig7} for three different system sizes. The same general behavior is observed for all of them: there is a maximum in the response of the system to the weak information input as a function of the idiosyncratic switching parameter $a$. As said before, this behavior is reminiscent of a well-known phenomenon generally labeled as resonance \cite{Benzi1981, Benzi1982, Nicolis1981, Collins1995, Gammaitoni1998}. The particular mechanism described here can be classified as an \emph{aperiodic stochastic resonance} \cite{Collins1996, Heneghan1996}, since the maximum in the response of the system to the external aperiodic signal is related to the relative importance of the stochastic term as compared to the deterministic one: the ratio between $h_0$ and $a$, in our case.

Within a financial market framework like the one presented in second part of the Methods section, a maximum in the response of the system would simply mean that the market optimally reproduces the level of optimism/pessimism contained in the incoming information (see, for example, panel~\textbf{b} of Fig.~\ref{fig6}). These results suggest that there is a certain range of idiosyncratic behavior intensity which largely improves the entrainment of the market by the external information. Note that the particular values of the response maxima are quite large, all of them implying a rather good fit of the information input. Note as well that these high levels of entrainment occur for a fairly wide range of values of the control parameter $a$: the input-output correlation stays above $0.7$ for more than a decade (in particular, for $h_0 \lesssim a \lesssim 20h_0$, with $h_0 = 10^{-3}$). However, there are two regions for which the system seems unable to follow the shape of the input signal: one for values of the idiosyncratic parameter smaller than the herding coefficient, and the other one for values much larger than the herding coefficient. In order to understand the differences between these two low-entrainment regimes, we also show in Fig.~\ref{fig7} a measure of the fraction of time spent by the system near the extremes of the opinion index space, $\sqrt{\langle x^2 \rangle}$. By observing this measure, it becomes clear that in the low-idiosyncrasy regime ($a < h_0$) the market cannot follow the input signal because the noise traders group amplifies the incoming information up to a complete consensus ($\sqrt{\langle x^2 \rangle} \simeq 1$) and then this group finds it difficult to leave this consensus and adapt to an updated information in the opposite sense (see also panel~\textbf{a} of Fig.~\ref{fig6}). On the contrary, in the high-idiosyncrasy regime ($a > 20h_0$) the problem faced by the market when trying to assimilate the incoming information is that, being the noise traders mostly guided by their own idiosyncratic drives, they pay little attention to external sources or other traders' attitudes, thus statistically tending to be equally divided among the two possible opinions ($\sqrt{\langle x^2 \rangle} \simeq 0$) (see also panel~\textbf{c} of Fig.~\ref{fig6}).


Bearing in mind that social and economic systems may have rather reduced sizes, it becomes relevant to assess the importance of size effects \cite{Toral2007}. In our case, the input-output correlation curves for the three different size examples in Fig.~\ref{fig7} collapse in the same curve for small values of $a$, while they are clearly different for intermediate and large values. This behavior can be understood in view of the functional form of the granularity term in the Langevin equation~\eqref{Langevin2}, the only one dependent on $N$: directly proportional to the idiosyncratic coefficient $a$ and inversely proportional to the system size $N$. For small values of the control parameter $a$, the granularity term is also small and the opinion index $x$ stays most of the time around the consensus states. In the intermediate $a$ region, the granular $N$ dependent term plays a relevant role in taking the system out of the consensus states with a probability uncorrelated with the shape of the input signal. Therefore, a large granularity term ---small number of agents $N$--- leads to a worse coincidence of $x$ with the input signal and, thereby, to a smaller input-output correlation. Finally, for large values of the control parameter $a$, the system does not even reach the extremes of the opinion index space and the behavior is predominantly led by the first term of the drift function and the noise effects produced by the granular term, because of the large value of $a$ appearing in both terms. Note as well that the relative difference between the curves is smaller when comparing the $N = 800$ with the $N = 200$ cases than when comparing this latter with the $N = 50$ example. This is simply a consequence of the granularity term being inversely proportional to the system size $N$, so for larger and larger $N$ the results become more and more similar.


\section*{Conclusions}
\label{sec:conclusions}


The aim of the present paper is to advance towards a quantitative understanding of the influence of an external source of information upon a financial market characterized by a certain herding behavior. The stochastic formalism used as a point of departure for our investigation incorporates individual behavioral heterogeneity as well as a tendency for social interaction, in the tradition of Kirman's seminal ant colony model. As opposed to the previous literature, which considers only the particular case of a closed market, we take into account the arrival of external information in the form of a time-dependent modification of the transition rates defining the individual traders' behavior.


A transition takes place in the original herding model from a monostable to a bistable behavior when increasing the herding propensity of the agents with respect to their idiosyncratic tendency. The monostable case can be understood as a market where each of two possible strategies is used by approximately half of the traders, while the bistable configuration corresponds to a market where there is always a clear majority of traders using one of the strategies, even if the chosen one can change over time. We have reinterpreted this \mbox{noise-induced} transition in terms of the mono- or bistability of an effective potential. In this context, we have demonstrated that the introduction of a dynamic external information input produces a time-dependent modification of this effective potential, whose symmetry is broken. We have used an \emph{Indicator of Economic Sentiment} published in Germany as an example of information input. Extensive simulations of this market model open to the arrival of external information have shown that even a small strength or convincing power of the external source may be enough for the market to follow its information signals. On the contrary, strong intensities lead to an amplification of the input signal: the convincing power of the external source being so strong, most of the traders are quickly persuaded to align their strategies in the sense of the input signal, giving rise to important market movements when the direction of this external information changes. Moreover, we have compared the results of this market model with Germany's leading stock market index, the DAX, showing that the introduction of a small strength information signal is able to reproduce general statistical properties of real financial data. In particular, the introduction of a low intensity signal allows the model to mimic: the volatility clustering effect, the slow decay of the autocorrelation of the normalized daily volatility for short time lags, and its zero and slightly negative values for long time lags.


Furthermore, we have studied the conditions for the market to show an optimal response to the arrival of external news, i.e., for it to optimally reflect the level of optimism/pessimism contained in the information input. The specific range of values for which an optimal response is observed depends on the intensity and the frequency or rate of change of the incoming information. In particular, we have found a certain range of values of the idiosyncratic behavior relative to the herding tendency among noise traders ($h_0 \lesssim a \lesssim 20h_0$) that optimizes their response as a group to a weak information input. We have shown the similarities of this phenomenon with an \emph{aperiodic stochastic resonance}. As a result of this analysis, we have identified three different market regimes regarding the assimilation of incoming information:
\begin{enumerate}
	\item Amplification of incoming information; any positive (negative) piece of news leads to a rather stable optimistic (pessimistic) consensus and it takes a long time for the market to adapt to updates of the sense of the external information (values of the idiosyncratic switching tendency below the range of the resonance, $a < h_0$).
	\item Precise assimilation of incoming information; the market optimally reflects the arrival of news (values of $a$ within the range of the resonance, $h_0 \lesssim a \lesssim 20h_0$).
	\item Undervaluation of incoming information; the arrival of news has an almost negligible influence and the market seems to be unaware of it (values of $a$ above the range of the resonance, $a > 20h_0$).
\end{enumerate}
A possible understanding of the origin of amplification in markets where traders are easily influenced by their peers ---markets dominated by collective herding behavior--- (regime 1) is that, once the external source of information is able to convince a small number of traders, they quickly spread the information to the rest of the market by influencing the decisions of other traders. On the contrary, in markets where investors behave independently of each other using their own expertise ---markets dominated by idiosyncratic behavior--- (regime 3), even if the external source is able to convince some of them, the information is not transmitted to the rest of the market: in order to convince the whole of the market, the external source would need to individually persuade each and every one of the traders. A precise assimilation of incoming information occurs when there is a compromise between these two factors (regime 2), i.e., when there is enough communication between traders and collective herding behavior to allow for the spreading of the external information to most of the market but also individual and independent behavior enough to prevent a full consensus in line with the external source of information.


The overreaction to incoming information can lead to explosions of fear or confidence triggered from outside the market, and thus not necessarily related to real changes in the fundamental value of the traded asset. We have shown that this amplification of external information is associated with the existence of short periods of enormous price variations, large volatility and, therefore, to a great instability of the market. The results presented in this paper support the idea that the greater importance of the herding with respect to the idiosyncratic tendency may play an important role in the development of such instabilities in a financial market open to the arrival of external information.


The analysis presented here considers only the quality of the response of the market to the arrival of external information. A natural step further would be to study the delay of the market in following the arrival of news, i.e., to take into account the time lag which leads to the maximum of the cross-correlation function between the input signal and the system output. Our results showing the existence of different market regimes regarding the assimilation of incoming news open the door for more comprehensive studies taking into account important features of today's information processing by market agents. A most relevant feature of real markets is the existence of different categories of investors, which could be characterized by different levels of sensitivity to incoming information, as empirically found in \cite{Lillo2015}. A further aspect to be considered is the asymmetric behavior that traders may have towards positive and negative news, leading, for instance, to explosions of fear but only slow waves of confidence. Note that asymmetry could also be introduced as a different sensitivity to news when prices are rising and when they are falling. Furthermore, we have focused here on the case of a global and passive reception of news. Thus, the modeling of an individual and active search for information, as empirically analyzed in \cite{Preis2010, Preis2013, Moat2013, Curme2014}, is left for future studies.

\newpage
\section*{Supporting Information}

\subsection*{S1 Appendix: Effective potential derivation}
\label{S1_Appendix}
\renewcommand{\theequation}{S1.\arabic{equation}}

We derive, in this appendix, the effective potential both for the original Kirman dynamics and the model with external information, presented in Eq.~$(8)$ and Eq.~$(20)$ in the main text. Let us start by restating here the definition of effective potential $U_{\textrm{eff}}(x)$ given in Eq.~$(7)$,
   \begin{equation}
   P_{\textrm{st}}(x) \equiv \mathcal{C}^{-1} \exp \left(-\frac{U_{\textrm{eff}}(x)}{D} \right),
   \end{equation}
where $P_{\textrm{st}}(x)$ is the stationary state probability distribution, $D$ is an effective noise intensity that we take as $D = h$, and the constant $\mathcal{C}^{-1}$ plays the role of a normalization factor. Note that, defined as such, the minima of this effective potential function will be attractive points of the dynamics, corresponding to maxima of the stationary state probability distribution.

For the general Fokker-Planck equation
   \begin{equation}
   \frac{\partial P_{\textrm{st}}(x,t)}{\partial t} = -\frac{\partial}{\partial x} \left[ q(x) P(x,t) \right] + \frac{\partial^2}{\partial x^2} \left[ D g(x)^2 P(x,t) \right],
   \end{equation}
the stationary distribution is found by assuming $\partial P_{\textrm{st}}/\partial t = 0$ and solving the resulting equation. By this means, a general effective potential \cite{SanMiguel2000} can be written as
   \begin{equation}
   U_{\textrm{eff}}(x) = -\int{\frac{q(x)}{g(x)^2}dx} + D \int{\frac{\partial g(x)}{\partial x} \frac{1}{g(x)}dx},
   \label{generalEffPot}
   \end{equation}
and, applying this definition to the Fokker-Planck equation~$(2)$ in the main text, the particular effective potential for the Kirman dynamics is found to be
   \begin{equation}
   U_{\textrm{eff}}(x) = (h - a)\,\ln(1-x^2).
   \end{equation}
Note that this effective potential $U_{\textrm{eff}}(x)$ is not to be confused with the deterministic potential, which is always monostable and can be found by simply integrating with respect to $x$ the deterministic part of Eq.~$(3)$.

Even though in the case with an external time varying forcing it is not possible to write a stationary state probability distribution, we assume that, at any point in time, the decay of the system to a quasi-stationary state is faster than the variation of the input signal, i.e., we assume conditions of slow driving. Therefore, we keep the previous definition of the effective potential as an approximation to this time-dependent case,
   \begin{equation}
   P(x,t) \approx \mathcal{C}^{-1} \exp \left(-\frac{U_{\textrm{eff}}(x,t)}{D} \right).
   \end{equation}
Thereby, applying equation~\eqref{generalEffPot} to the Fokker-Planck equation~$(18)$ leads to the particular functional form
   \begin{equation}
   U_{\textrm{eff}}(x,t) = (h_0 - a)\,\ln(1-x^2) - x F i(t)
   \end{equation}
for the model with arrival of external information.

\newpage
\subsection*{S1 Figure: Probability distribution of absolute normalized daily returns}
\label{S1_Figure}

\begin{figure}[ht!]

	\centering

	\includegraphics[width=0.85\textwidth, height=!]{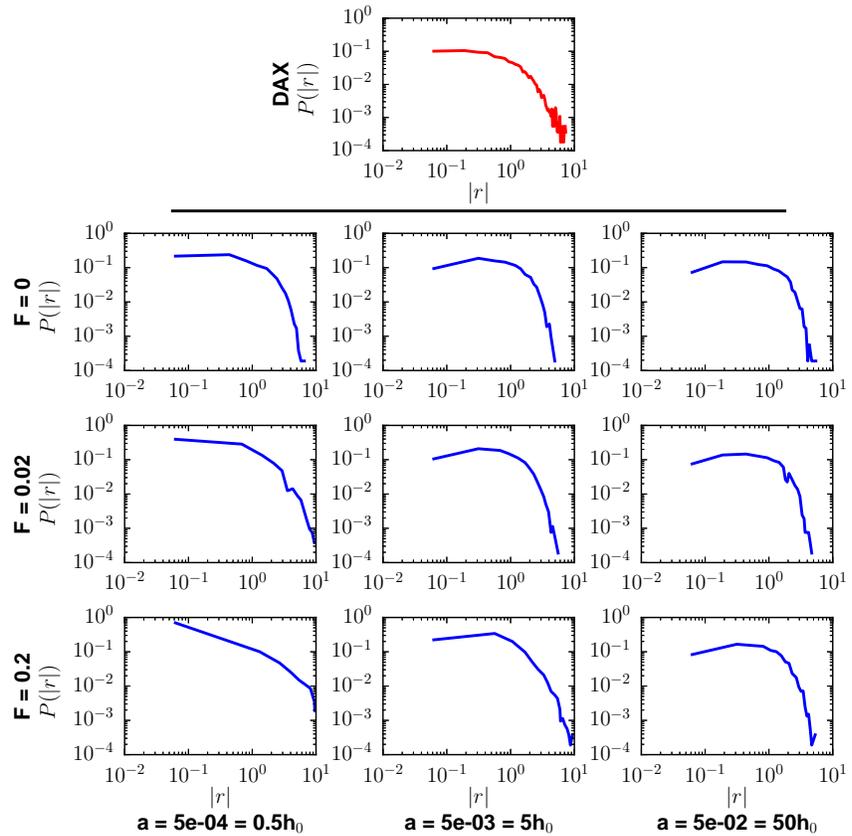}

	\caption{{\bf Probability distribution of the absolute value of the normalized daily returns, $P(|r|)$.} First panel, red line: German DAX index from December 1991 to November 2013, shown for comparison. Table of nine panels, blue lines: Model results for three different input intensities of the external information $F$ and three values of the idiosyncratic switching tendency $a$. The rest of the parameters are fixed as $h_0 = 10^{-3}$ and $N = 200$.}
\end{figure}


\section*{Acknowledgments}
We are grateful to Marco Patriarca for helpful comments and suggestions during the preparation of this paper.

\end{document}